# Real-time Adaptive and Localized Spatiotemporal Clutter Filtering for Ultrasound Small Vessel Imaging

Chengwu Huang[1], U-Wai Lok[1], Jingke Zhang[1], Hui Liu[1,2], Shigao Chen[1]

[1]Department of Radiology, Mayo Clinic College of Medicine and Science, Rochester, MN, USA

[2]Department of Ultrasound, The First Affiliated Hospital of Fujian Medical University, Fuzhou, China

Corresponding Author: Shigao Chen, Department of Radiology, Mayo Clinic, 200 First Street SW, Rochester, MN 55905, USA Chen.Shigao@mayo.edu

**Funding:** The study was partially supported by the National Institute of Diabetes and Digestive and Kidney Diseases under award number R01DK120559 and R01DK129205, and by National Institute of Arthritis and Musculoskeletal and Skin Diseases under award number R21AR076028. The content is solely the responsibility of the authors and does not necessarily represent the official views of the National Institutes of Health.

**Competing Interests**: The Mayo Clinic and some of the authors (C.H. and S.C.) have pending patent application related to the technology referenced in the research.

ABSTRACT

Effective clutter filtering is crucial in suppressing tissue clutter and extracting blood flow signal in Doppler ultrasound. Recent advances in eigen-based clutter filtering techniques have enabled ultrasound imaging of microvasculature without the need for contrast agents. However, simultaneously achieving fully adaptive, highly sensitive and real-time implementation of such eigen-based filtering techniques in clinical scanning scenarios for broad translation remains challenging. To address this, here we propose a fast spatiotemporal clutter filtering technique based on eigenvalue decomposition (EVD) and a novel localized data processing framework for robust and high-definition ultrasound imaging of blood flow. Unlike the existing local clutter filter that hard splits the ultrasound data into small blocks, our approach applies a series of 2D spatial Gaussian windows to the original data to generate local data subsets. This approach improves performance of flow detection while effectively avoiding undesired grid artifacts with dramatically reduced number of subsets required in local EVD filtering to shorten computation time. By leveraging the computational power of Graphics Processing Units (GPUs), we demonstrate the real-time implementation capability of the proposed approach. We also introduce and systematically evaluate several adaptive and automatic eigenvalue thresholding methods tailored for EVD-based filtering to facilitate optimization of blood flow imaging for either global or localized processing. Besides the widely adopted hard thresholding, soft eigenvalue thresholding methods by assigning non-binary weights to eigen components are also evaluated for their efficacy in blood flow imaging. The feasibility of the proposed clutter filtering technique is validated by experimental results from phantom and different *in vivo* studies, revealing robust clinical application potential. A tradeoff between improved performance and computational cost associated with the packet size and subset number in local processing is also investigated.

INTRODUCTION

Doppler ultrasound is widely used in clinical settings for imaging and quantification of blood flow. Clutter filtering is a key signal processing procedure in Doppler ultrasound to suppress unwanted tissue signal (*i.e.*, tissue clutter) and extract flow signal for subsequent imaging and measurement [1]. Blood flow signal, however, is commonly significantly weaker in magnitude compared to the surrounding tissue clutter at clinical ultrasound frequencies [2]. Hence, development of a highly sensitive clutter filter is essential to improve blood flow detection, especially for slow blood flow in small vessels. Assuming that blood flow exhibits a higher temporal frequency compared to tissue motion, conventional clutter filters primarily rely on a high-pass filtering along temporal dimension (*i.e.*, slow-time dimension) to suppress tissue clutter [1, 3]. Nevertheless, high-pass filtering is known to have limited sensitivity to slow blood flow [4-7]. Additionally, the presence of tissue motion can lead to increased clutter bandwidth and frequency shifts in the Doppler spectrum, resulting in overlap with the blood flow signal, and weakening the capability of conventional high-pass filters to separate blood flow and tissue clutter [4]. More advanced clutter filters based eigen-value decomposition (EVD), or singular value decomposition (SVD), have been explored in Doppler ultrasound that enable improved performance in blood flow detection. Pioneering studies in the field have been conducted by various research groups [8-16] and extensively reviewed by Yu and Lovstakken [4]. These applications of eigen-based filters, however, primarily lies within the context of the conventional focus line-by-line ultrasound scanning strategies. In recent years, the advent of ultrafast ultrasound imaging that based on transmission of unfocused beams such as plane wave or diverging wave, has enabled wide-field ultrasound imaging at extremely high frame rate [17, 18]. Combining ultrafast imaging with SVD-based filtering that leverages both long temporal ensemble length (packet size) and the rich spatial information has resulted in substantially improved discrimination of blood flow and tissue clutters [5-7]. These techniques have demonstrated superior Doppler sensitivity to small vessels compared to conventional Doppler ultrasound, fostering a diverse applications, including functional ultrasound imaging of brain and spine, tumor evaluation, and assessment of kidney, liver, inflammatory diseases [5, 19-22].

The general principle underlying these eigen-based clutter filters involves projecting the original data onto a new set of bases that enables better discrimination of tissue clutter and blood flow components. Filtered data can be reconstructed through transforming a selected subset of components corresponding to blood flow back to the image domain [4, 23-25]. Subsequently, two important technical considerations arise: 1) the ability to effectively separate tissue and blood components in the transformed domain, and 2) the selection of appropriate components in the transformed domain that represent blood flow signals. To improve ability of clutter/blood separation, in additional to the commonly used SVD filters, several other blind source separation methods such as independent component analysis (ICA) have been explored [23, 26]. High-order SVD method that expands the dimension of the data in SVD analysis have also demonstrated improved separation performance [27-29]. Low-rank and sparse matrix decomposition methods like robust principal component analysis (RPCA) have also been attempted and shown promise for improved clutter filtering performance by assuming the low-rank structure of tissue clutter and spatial-temporally sparsity of blood flow signal [30-33]. However, these methods often impose a significantly larger computational requirements that may limit their real-time implementation. Most of the current methods rely on performing SVD analysis on full field-of-view (FOV) ultrasound data, assuming consistent tissue and noise characteristics across the entire spatial FOV. This assumption may be violated in complex clinical scanning scenarios due to spatially varying clutter, noise, and tissue motion characteristics. To address this issue, a block-wise SVD clutter filtering technique was previously proposed and demonstrated superior performance in small vessel imaging [34]. This approach involves dividing the full FOV data into overlapping local blocks and processing each block separately based on local data statistics. A significant drawback of the block-wise SVD method, however, is the high computational cost associated with the large number of SVD calculations needed for the numerous data subsets. Although accelerated SVD processing techniques like randomized SVD (rSVD) [35, 36] or simplified SVD method [37] can be employed to expedite the calculations, the large number of data subsets required in block-wise processing still compromise the fast implementation. Additionally, "grid pattern" artifacts resulting from the energy discontinuities between adjacent filtered blocks, particularly when the block overlapping ratio is low, can severely degrade overall image quality [34].

An accurate and adaptive selection of appropriate components representing blood flow signals and tissue clutter in the eigen domain is another critical aspect of clutter filtering. This becomes particularly important in local SVD filtering, as variations of singular value thresholding can lead to significant energy difference among neighboring blocks, worsening the grid artifact in the resulting blood flow image [34]. Various adaptive criteria have been proposed to identify thresholds that optimally separate the tissue and blood flow components. These criteria leverage characteristics such as eigen or singular values, temporal spectral content, and spatial content or a combination [4, 23, 24, 28, 34, 38-40]. For instance, assuming distinct signal energy levels for clutter and blood flow, an appropriate threshold can be determined by analyzing the distribution of eigenvalues or singular values, such as thresholding based on predefined magnitude, difference or ratio of the eigenvalues or singular values [4, 23, 24, 38]. The "turning point" of the sorted singular value curve, defined by a given slope or peak acceleration, has also demonstrated effective as a cutoff to discriminate between clutter and flow [23, 34, 38, 39]. Frequency analysis of each eigen vector can also guide threshold selection considering the differences in temporal characteristics of tissue and blood flow [4]. Recent advancement has leveraged spatial similarity among different singular components to improve discrimination, as tissue, blood flow, and noise exhibit inherently distinct spatial patterns [38, 41, 42]. More comprehensive reviews of the different criteria and principles can be found in studies by Wildeboer et al. and Baranger et al. [23, 38]. However, achieving fully automatic component selection remains challenging for real-time clinical application, especially when incorporating localized data processing. Most current adaptive methods may still require a predefined value to initiate the automatic calculation of thresholds. For example, a proper predefined ratio or slope of singular value or a predefined center frequency of singular vector will be needed in the calculation of finding the singular value thresholds [23, 34, 38]. Different predefined values may be required to achieve optimal clutter suppression in different applications, typically determined and modified empirically on a case-by-case basis. Additionally, while most methods employ hard thresholding of singular values, where eigen components are either completely removed or entirely preserved, the soft thresholding strategy that assigns non-binary weights to each of the eigen components has not been widely explored in ultrasound microvessel imaging [23, 24, 43].

Based upon the current technical developments, this study proposes a robust and novel localized clutter filtering framework based on EVD for high-sensitive, fully adaptive, and fast implementation of blood flow imaging. Instead of hard splitting the data into small blocks, the proposed technique addresses the limitations of current block-wise SVD filter by applying a series of 2D Gaussian windows to the original data to generate data subsets. This approach effectively avoids grid artifacts and the number of data subsets in local EVD filtering can thus be dramatically reduced to shorten processing time for real-time implementation. We also introduce and systematically compare several new adaptive and automatic thresholding methods for eigen component selection to facilitate robust and efficient implementation of the localized EVD filtering. Notably, besides the widely adopted hard thresholding method for component selection, soft thresholding methods that assign non-binary weights to the eigen components are also introduced and evaluated for their efficacy in blood flow imaging. The proposed technique was successfully implemented in real-time facilitated by Graphics Processing Units (GPUs) on various realistic clinical datasets, demonstrating its great potential for successful clinical translation.

MATERIALS AND METHODS

*Principles of EVD-based Clutter Filtering*

In Doppler ultrasound, clutter filtering is commonly applied to a packet of ultrasound data acquired at a specific pulse-repetition frequency (PRF). For conventional Doppler ultrasound, a small packet size (also known as ensemble size) of approximately 8 to 24 frames is often used for tissue clutter filtering. However, in the context of ultrafast Doppler processing, a larger packet size of ≥50 frames is typically recommended to enhance the performance of eigen-based spatial-temporal clutter filtering [36]. Consider a packet of the acquired ultrasound data, **S**, to have a dimension of $N_y \times N_x \times N_t$, where $N_x$ and $N_y$ correspond to lateral and axial spatial dimensions respectively, and $N_t$ represents the temporal dimension (*i.e.*, packet size). The data **S** comprises different components, mainly including tissue clutter, blood flow signal and noise. Prior to spatial-temporal clutter filtering to suppressing tissue clutter, the 3D matrix **S** is typically reshaped into a 2D Casorati matrix (**C**) with a dimension of $N_x N_y \times N_t$, as illustrated in Fig. 1, where the columns of the matrix **C** represent vectorized frames [5].

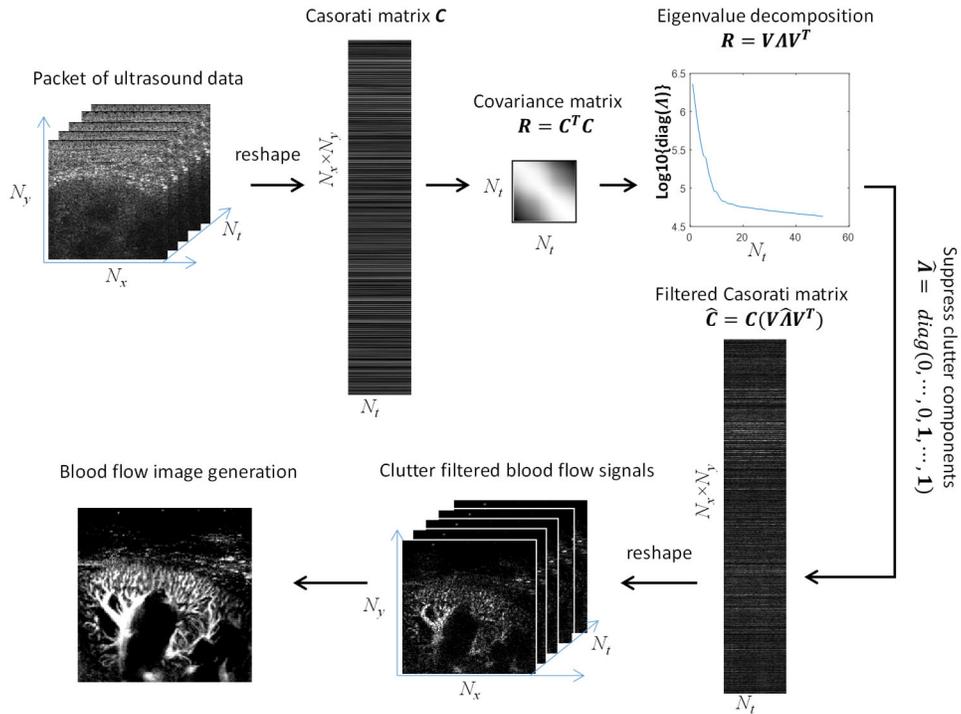

Fig. 1 Illustration of EVD-based tissue clutter filtering for blood flow imaging

The EVD-based clutter filtering starts by computing a covariance matrix of the $C$, as:

$$R = C^T C = \sum_{i=1}^{NxNy}(s_i^T s_i) \tag{1}$$

where $s$ is the row vector of $C$ (*i.e.*, a data vector extracted from each spatial position with a size of 1×*Nt*). By calculating the EVD of the covariance matrix $R$, the eigenvalues and eigenvectors can be obtained as:

$$R = V \Lambda V^T \tag{2}$$

where $V = [v_1, v_2, \cdots, v_{Nt}]$ is an *Nt*×*Nt* orthonormal matrix whose columns are the eigenvectors, and $\Lambda = diag(\lambda_1, \lambda_2, \cdots, \lambda_{Nt})$ is an *Nt*×*Nt* diagonal matrix of the eigenvalues [4, 24, 44]. The eigenvalues in $\Lambda$ are assumed to be sorted in descending order in magnitude. The computation of eigenvalues and eigenvectors for the covariance matrix $R$ is in theory equivalent to the computation of the SVD of the Casorati matrix $C$ in this case. For the SVD definition $C = UDV^T$, we have:

$$C^T C = (UDV^T)^T UDV^T = V(D^T D)V^T \tag{3}$$

where $U$ is an orthonormal matrix whose columns are the left singular vectors of $C$, denoted $U = [u_1, u_2, \cdots, u_{Nt}]$; and $V$ is the orthonormal matrix whose columns are the right singular vectors, which is identical to the eigenvectors; and $D$ is a diagonal matrix of the singular values of $C$, expressed as $D = diag(\sigma_1, \sigma_2, \cdots, \sigma_{Nt})$, with singular values sorted in descending order. Therefore, according to Eqns. 2 and 3, the eigenvalues of $R$ is equivalent to the square of the corresponding singular value of $C$, *i.e.*, $\lambda_i = \sigma_i^2$. Hence, like the SVD-based clutter filtering, tissue clutter is represented by the low-order large eigenvalues in EVD-based method, while the blood flow signals are represented by the median to high-order smaller eigenvalues and the corresponding eigenvectors. By rejecting the components associated with tissue clutter, the blood flow signals can be recovered by a projection procedure [4, 24, 44]:

$$\hat{C} = C(V\hat{\Lambda}V^T) \tag{4}$$

where $\widehat{C}$ is the obtained blood flow signal with tissue clutter suppressed, and $\widehat{\Lambda} = diag(0,\cdots,0,1,\cdots,1)$, which is the diagonal matrix of the ones but rejecting the first $k^{th}$ diagonal entries associated with tissue clutter by setting them to zeros in the hard eigenvalue thresholding methods. Several adaptive methods for selecting optimal number eigen components, $k$, to be suppressed will be detailed in the next section. $\widehat{C}$ can then be reshaped back to the original data dimension of $Nx \times Ny \times Nt$ for generation of blood flow image (Fig. 1). As an alternative, the tissue clutter filtering can be achieved by first reconstructing the principal components representing the tissue clutters and then subtracting them from the original data, as:

$$\widehat{C} = C(I - V\breve{\Lambda}V^T) \tag{5}$$

where $I$ is the identity matrix, and $\breve{\Lambda} = diag(1,\cdots,1,0,\cdots,0)$, which only preserves the first $k^{th}$ eigenvalues representing tissue components. From the clutter filtered signals, power Doppler image can be estimated as the mean or summation of the signal power along temporal dimension [5, 7].

In the context of ultrafast Doppler processing, the Casorati matrix often has a much larger number of rows than columns, especially for real-time applications when a relatively small packet size (ensemble size) is allowed (*e.g.*, packet size of 50) [36]. For instance, in the case of an ultrasound Doppler data with a spatial-temporal dimension of 100×100×50 (*Nx×Ny×Nt*), the reshaped Casorati matrix has 10000 rows, which is 200 time larger than the number of columns. In the principal component analysis employed in this study, the calculation of covariance matrix (Eqn. 1) is a matrix multiplication operation of 'tall' matrices which can be efficiently accelerated by leveraging the massive parallelization of GPUs [45, 46]. The resultant size of the covariance matrix is much smaller (*Nt×Nt*, solely determined by the packet size, as example shown in Fig. 1), which enables an efficient calculation of EVD, given the EVD computational complexity of $O(Nt^3)$ [47].

*Adaptive Thresholding Methods for EVD-based Clutter Filtering*

In the pursuit of real-time implementation of eigen-based clutter filtering, a fully adaptive, automatic, and computationally efficient method for identifying tissue clutter components is crucial. Tissue clutter is typically characterized by low-order eigenvalues with larger signal energy and higher spatial-temporal coherence, while blood signal is represented by middle-to-small eigenvalues [34]. The primary goal is to accurately identify the eigen components with the largest eigenvalues (or the number of low ranks) that correspond to tissue clutter and suppress them from the data. In this study, we introduce or adopt several methods and will comprehensively evaluate their performance in various applications.

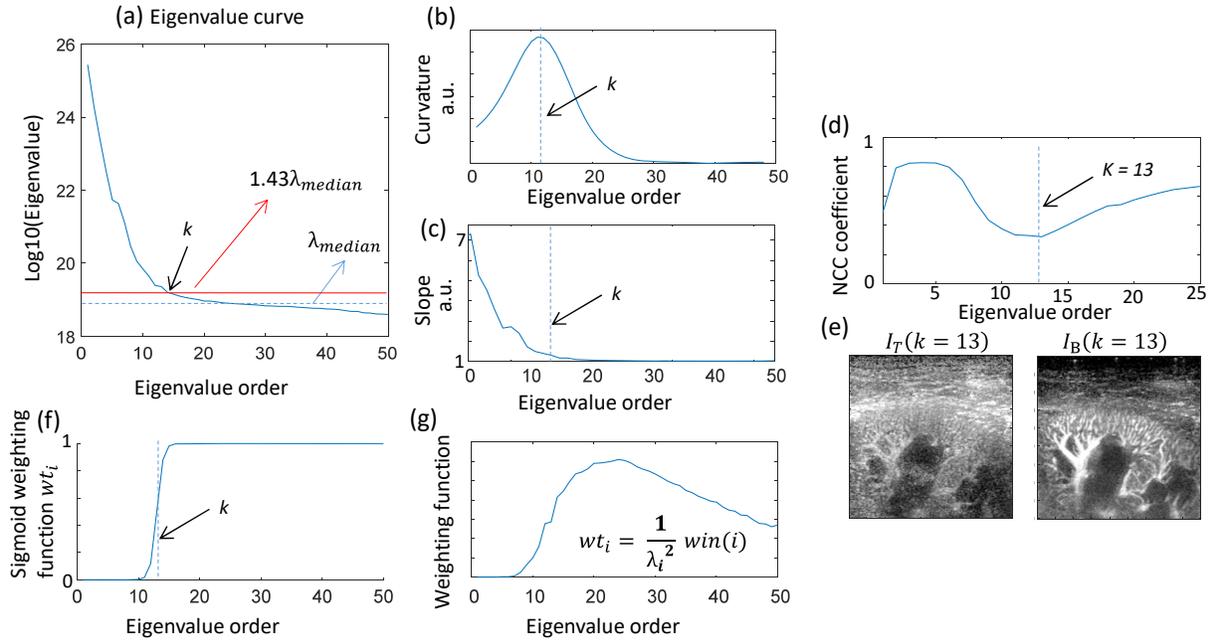

Fig. 2. (a) An example eigenvalue curve. An eigenvalue threshold can be efficiently and adaptively determined by the scaled median eigenvalue, as indicated by the red line (here scaling factor = 1.43). (b) Curvature of the eigenvalue curve, maximum of which gives the location of eigen cutoff ($k$). (c) The slope of the eigenvalue curve. A predefined slope threshold can be used to identify the cutoff $k$. (d) The normalized cross-correlation (NCC) coefficient of the tissue image $I_T(k)$ and blood image $I_T(k)$ as a function of eigenvalue cutoff ($k$). The optimal $k$ is identified at the location with a minimum NCC coefficient ($k$=13 in this case), where the reconstructed $I_T(k)$ and $I_B(k)$ are assumed to be most distinct, as exemplified in (e). (f) Sigmoid weighting function for a soft transition between tissue and blood eigenvalues. (g) Adaptive weighting function based on the eigenvalue magnitude, to be applied to the eigenvalue curve for soft eigenvalue thresholding.

1. Eigenvalue thresholding based on maximizing the curvature of the eigenvalue curve.

The transition from tissue components to blood components is often identifiable as a 'turning point' on the eigenvalue curve, as depicted in Fig. 2a, where the eigenvalues start to flatten. This transition location can be characterized as a position with the maximum local curvature. Here, considering the eigenvalue as a function of eigenvalue order $n$ in logarithmic scale, $y(n) = \log_{10} \lambda_n$, the signed curvature can be estimated as:

$$\mathcal{K} = \frac{|y''|}{(1+y'^2)^{3/2}} \tag{6}$$

where $y'$ and $y''$ are the first and second derivatives of $y$, respectively. In practice To refine the determination of the transition location and allow fine-tuning of the thresholding, the curve is normalized as $\beta \frac{y(n)}{y(1)} Nt$ before computing the curvature, where $y(1)$ is the largest eigenvalue, $Nt$ is number of eigenvalues (packet size) and $\beta$ is a predefined scaling factor that can slightly shift the location of curvature maxima for fine-tuning. The optimal $\beta$ may vary depend on the specific data and application. Figure. 2b shows an example of the curvature estimation with $\beta = 1$. The maximum curvature gives the estimate of threshold cutoff (denoted as $k$), which is used as the number of eigen components to be rejected. We will assess the method with varying scaling factor $\beta$ to optimize eigenvalue thresholding for different applications. As curvature is the reciprocal of the local radius of a curve, finding the maximum curvature is also equivalent to finding the minimum radius of the curve.

2. Eigenvalue thresholding based on slope of the eigenvalue curve.

Finding the transition location or the 'turning point' on singular value curves through slope analysis has been previously demonstrated as a feasible approach [34, 38]. This method calculates the slope, or the first derivative of the singular value curve and requires a properly predefined slope threshold above which the singular values are removed. The predefined slope threshold is empirically determined and can be data-dependent for optimal tissue suppression. Figure. 2c shows an example of the slope estimation of the eigenvalue curve $y(n) = \lambda_n$ and the identified eigenvalue cutoff based a predefined slope threshold of

1.1. The influence of varying slope threshold on clutter filtering across various applications will be evaluated and used as a benchmark for investigating other methods.

3. Eigenvalue thresholding based on the median eigenvalue.

An optimal hard threshold in SVD can be theoretically derived for matrix denoising under the assumption that a matrix has a low-rank structure contaminated with Gaussian white noise. This theoretical model establishes a simple relationship between the median singular value and optimal hard threshold when the data noise level is unknown [48]. For instance, the optimal threshold is approximately $2.858\sigma_{median}$ for the recovery of low rank square matrices, while ratio of the optimal threshold and $\sigma_{median}$ can vary and be numerically determined for rectangular matrices, with $\sigma_{median}$ being the median of the singular values [48]. While this theoretical model may not be directly applicable to the Doppler clutter filtering applications, as the flow signal is not completely identical to random noise, the concept of linking optimal threshold with the median singular value can be leveraged. For simplicity, we assume that the optimal threshold can be proportional to median eigenvalue in the case of EVD, as:

$$\mathcal{T} = \beta \cdot \lambda_{median} \qquad (7)$$

Where β is a scaling factor, and $\lambda_{median}$ is the median eigenvalue. The eigen components with eigenvalues larger than this threshold $\mathcal{T}$ can be considered as the low-rank tissue clutter and can thus be removed. Once a scaling factor β is provided, the optimal threshold can be efficiently and adaptively determined. Figure. 2a illustrates an example of the identified threshold given β = 1.43. This study investigates the effectiveness of this method across various applications to guide the determination of appropriate scaling factors.

4. Eigenvalue thresholding based on minimizing the spatial correlation of tissue and blood flow components.

Inspired by the studies by Baranger *et al* and Kim *et al* [28, 38], the method presented here is based on the assumption that at an optimal threshold, the reconstructed blood flow signal and the tissue clutter should exhibit the most distinct spatial patterns (*i.e.*, minimized spatial correlation). Ideally, for an optimal eigenvalue cutoff ($k$), the tissue image (reconstructed from the first $k$ eigen components) and blood flow image (reconstructed from the rest of the eigen components) should have the most distinct spatial patterns: blood flow image reveals the vascular morphology, while tissue image shows the tissue-related structure. The problem here can be modeled as finding a $k$ that minimize the normalized spatial cross-correlation coefficient between tissue clutter image and the blood flow image:

$$\arg\min_{k} \left\{ \frac{\sum I_T(k) \cdot I_B(k)}{\sqrt{\sum I_T(k)^2} \cdot \sqrt{\sum I_B(k)^2}} \right\}, \; for \; k = 1, 2, \ldots Nt \tag{8}$$

Where $k$ is the threshold cutoff that identifies the first $k^{th}$ number of eigen components as tissue clutter, the rest of the eigen components as blood flow and noise. $I_T(k)$ and $I_B(k)$ are the reconstructed tissue and blood flow images respectively, at a given $k$.

Directly reconstructing a tissue image and a blood flow image for every $k$ (Eqn. 4) involves a substantial computational cost. A more efficient approach is to leverage the left singular matrix $\boldsymbol{U} = [\boldsymbol{u_1}, \boldsymbol{u_2}, \cdots, \boldsymbol{u_{Nt}}]$, where the columns of which are the left singular vectors that represents the spatial patten of each eigen components [28, 38]. $\boldsymbol{U}$ can be calculated from the eigenvectors according to $\boldsymbol{CV} = \boldsymbol{UD}$ and $\lambda_i = \sigma_i^2$. Therefore, for a given $k$, an alternate of the tissue image $I_T(k)$ can be obtained from the first $k^{th}$ left singular vectors $I_T(k) = \sum_{i=1}^{k} w_i |\boldsymbol{u_i}|$, and blood flow image $I_B(k)$ can be obtained from the rest of the left singular vectors $I_B(k) = \sum_{i=k+1}^{Nt} w_i |\boldsymbol{u_i}|$, where $|\boldsymbol{u_i}|$ indicates the magnitude of $\boldsymbol{u_i}$, and $w_i$ is a weight applied to each spatial singular vectors. To balance the number of singular vectors to generate $I_T(k)$ and $I_B(k)$, we further simplify the definition as $I_T(k) = \sum_{i=k-fn+1}^{k} w_i |\boldsymbol{u_i}|$ and $I_B(k) = \sum_{i=k+1}^{k+fn} w_i |\boldsymbol{u_i}|$ that considers a given number *fn* of left singular vectors next to the cutoff $k$. Here, we consider *fn* = 5, and $w_i = 1$ for consistent evaluation. By minimizing the correlation between $I_B(k)$ and

$I_T(k)$ based on Eqn. 8, an optimal $k$ that best separates the tissue clutter and blood flow signal is expected. This is efficiently achieved by finding the minima of the cross-correlation curve, without the need for calculation of the full spatial similarity matrix. In the example shown in Fig. 2d, the optimal $k$ is identified as 13 to reach a minimum correlation coefficient, where the $I_T(k)$ and $I_B(k)$ are considered most distinct (Fig. 2e).

5. Soft thresholding based on the sigmoid function.

In practice, complete separation of tissue clutter and blood flow in eigen domain is challenging, especially in the presence of tissue motion. Consequently, there tends to be some overlap between tissue clutter and blood flow components in the eigen spectrum, where rejected tissue eigen components may contain parts of flow signals in hard thresholding. Instead of hard thresholding that completely retain or reject eigen components, here we propose to apply soft thresholding strategy by applying a non-binary weight defined by a sigmoid function to the eigenvalues to enable a soft transition between tissue and blood flow subdomains [24]. The sigmoid function can be defined as:

$$wt_i = 1/(1 + e^{-\beta(i-k)}) \qquad (9)$$

where $\beta$ is a predefined scaling factor for the sigmoid function, which controls the sharpness of the transition, $i$ is the eigenvalue order ($i$ = 1 to $Nt$), $k$ is the eigenvalue cutoff provided by the hard thresholding methods. The eigenvalue cutoff ($k$) obtained from the aforementioned method 4 will be used here to comparatively evaluate this soft thresholding method with varying predefined scaling factor $\beta$. In this method, instead of setting the first $k^{th}$ eigenvalues to be zeros, the eigenvalues are weighted in diagonal matrix as $\widehat{\Lambda} = diag(wt_1, wt_2, \cdots, wt_{Nt})$, followed by Eqns. (4) or (5) to reconstruct filtered blood flow signal. Figure. 2f depicts an example of the sigmoid based weighting function ($\beta$ = 2, $k$ =13) to weight the eigenvalue curve for tissue clutter filtering.

6. Adaptive soft thresholding based on signal energy of eigen components.

Given that tissue clutter and blood flow components exhibit differences in their contributions to signal energy, we propose an adaptive soft thresholding method that allows for flexible design of eigen weights based on the eigenvalue magnitude. Specifically, clutter components, characterized by larger eigenvalues, can be assigned smaller weights to achieve tissue suppression, while flow components corresponding to smaller eigenvalues can be assigned larger weights for enhancement. As a simple example, weights can be inversely related to the eigenvalue magnitude, given by:

$$wt_i = 1/\lambda_i^{\beta} \qquad (10)$$

where β > 0 is a predefined value controlling the order of the weighting function. However, this weighting design may enlarge the undesirable electronic noise components corresponded to very small eigenvalues, compromising the SNR of flow detection. To mitigate this effect, here we apply a Gaussian window function to Eqn. 10 to attenuate the weights for the high-order eigenvalues, as $wt_i = win(i)/\lambda_i^{\beta}$, where *win*(*i*) represents the Gaussian window center at *i*=0 (generated by 'gausswin' function in Matlab). The weighting function $wt_i$ is normalized before assigning to eigenvalues for clutter filtering, and the performance of method will be evaluated with changing predefined factor β. Figure 2g shows an example of weighting function $wt_i$ (β=2) applied to the eigenvalue curve, indicating the suppression of larger eigenvalues (related to tissue clutter) and very small eigenvalues (related to noise), as well as the enhancement the median eigenvalue (mainly related to blood flow signal).

*Localized Clutter Filtering Via 2D Spatial Windowing*

The utilization of localized processing holds promise for enhancing tissue clutter filtering especially for realistic clinical data with spatial heterogeneity [34]. Previous block-wise local processing technique that divides the data into densely overlapping subsets suffers from limitations such as high computational cost and grid artifacts. Here in our method, instead of hard splitting the data into spatially overlapping blocks, we apply different 2D spatial weighting windows to the original ultrasound data separately, to generate multiple windowed ultrasound datasets, as exemplified in Fig. 3. Each windowed data subsets will undergo the EVD-based clutter filtering described above to generate filtered blood flow signal. Combination of the filtered blood flow signals from different subsets will lead to the final blood flow image with improved performance. As data energy is localized at the center of window for each subset, clutter filtering is expected to benefit from this localized data characteristic, while the smooth transition between data subsets can significantly mitigates energy discontinuities when combining the filtered subsets.

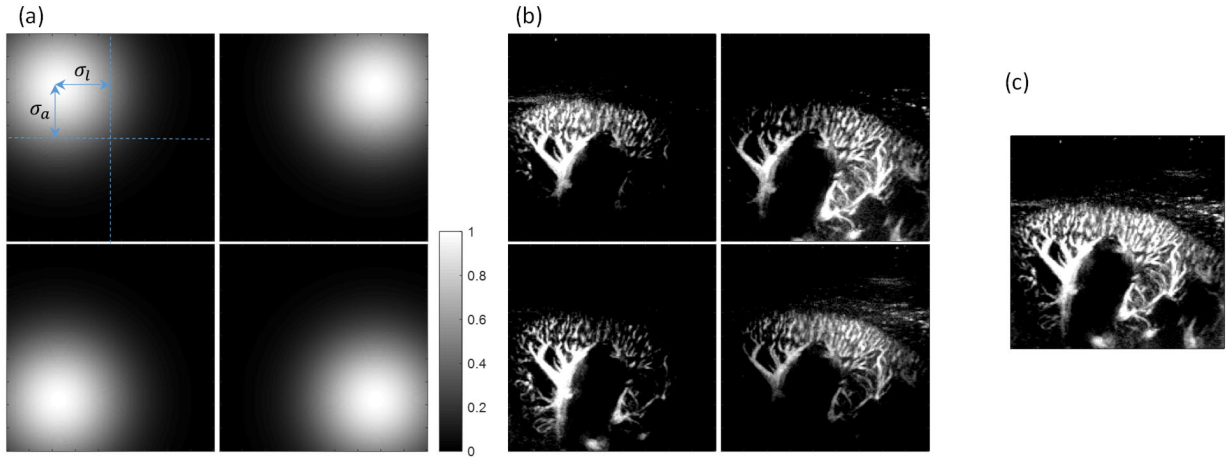

Fig. 3. Illustration of localized clutter filtering via spatial windowing. (a) An example of a set of four (2×2) Gaussian windows centered at four different positions, applied to the original ultrasound data separately to generate multiple windowed ultrasound datasets. (b) Corresponding blood flow images of these datasets obtained from tissue clutter filtering. (c) The final blood flow image obtained from the combination of different subsets.

As an illustrative example shown in Fig. 3a, a set of four Gaussian windows centered at four different positions are used to generate four subsets for clutter filtering. Assuming the Casorati matrix of the original ultrasound data is $\boldsymbol{C} = \begin{bmatrix} s_1 \\ \vdots \\ s_{N_xN_y} \end{bmatrix}$, where $s_i$ represents the signal at a spatial position with a size of 1×Nt, also referred to as slow-time signal at that position, and $\boldsymbol{w} = \begin{bmatrix} w_1 \\ \vdots \\ w_{N_xN_y} \end{bmatrix}$ is the N$x$N$y$×1 Casorati matrix of one of the 2D Gaussian windows (Fig. 3a), with $w_i$ being the weight at the same spatial position of $s_i$, the Casorati matrix of the windowed ultrasound data is now expressed as:

$$C_w = WC = \begin{bmatrix} w_1 \cdot s_1 \\ \vdots \\ w_i \cdot s_i \\ \vdots \\ w_{NxNy} \cdot s_{NxNy} \end{bmatrix} \quad (11)$$

Where $W$ is a diagonal matrix of $w$, i.e., $W = diag(w)$. The weighted ultrasound data thus represents the data information of the localized positions defined by the windows center where the ultrasound energy is dominant. The covariance matrix of the Casorati matrix $C_w$ can computed for each windowed data, as:

$$R = C_w^T C_w = C^T W^T W C = \sum_{i=1}^{xy} w_i^2 (s_i^T s_i) \quad (12)$$

Thus, $R$ is viewed as a weighted summation of $s_i^T s_i$, where the weights here are the square of the weights applied to the original ultrasound data, *i.e.*, $w_i^2$. Following EVD calculation of the $R$ and suppressing the eigen components corresponding to the tissue clutter using the proposed adaptive thresholding methods, the filtered blood flow signal can be reconstructed for each subset (Fig. 3b). The final power Doppler image of the blood flow is then generated by weighted sum of the individual power Doppler images from different subsets:

$$PD_{final} = \sum_{i=1}^{Nd} \alpha_i PD_i \quad (13)$$

Where $Nd$ is the total number data subsets, and $\alpha_i$ is the weight applied to the $i^{th}$ subsets. Here, the $\alpha_i$ is chosen as reciprocal of mean flow signal power, resulting in the final blood flow image as the summation of separated blood flow images normalized to their signal power (Fig. 3c). The window center positions, and the standard deviation (SD) of the Gaussian windows are adaptively determined by the number of data subsets used to cover the full FOV of the data, as indicated in Fig. 3. Given the smooth transition of the Gaussian windows, no abrupt change of blood flow energy is expected at the boundaries, and thus a large number of subsets, as required in block-wise SVD filtering [34], is not necessarily needed, facilitating the fast implementation of the method.

*Experimental Studies*

1. Flow phantom study

The performance of the proposed methods was initially evaluated using a customized flow channel phantom (Gammex Inc., Middleton, WI, USA) with a 2 mm inner diameter flow channel. The outlet of the flow channel was connected to a syringe pump (Model NE-1010, New Era Pump Systems Inc., Farmingdale, NY, USA) which provided steady flow of ultrasound blood mimicking fluid (Gammex Inc., Middleton, WI, USA) through the flow channel at a constant flow rate (1.89 ml/min, corresponding to an average flow speed of 10 mm/s). Ultrasound plane wave compounding imaging (10-angle plane wave compounding, with 1º increment between successive transmissions) was performed using a Verasonics Vantage system (Verasonics Inc., Kirkland, WA, USA) equipped with a GE 9L linear array probe (GE Healthcare, Wauwatosa, WI, USA). The imaging parameters included a center frequency 6.25 MHz, transmit pulse length of 2 cycles, and a frame rate of 500 Hz for the post-compounded data. A stack of 496 frames in-phase quadrature (IQ) data were captured from the longitudinal plane of the flow channel (data size of 196×180×496 with data type of complex double). The pixel size was set to be 0.246 mm in both axial and lateral directions. To mimic *in vivo* tissue motion, an external mechanical shaker (LDS Model V203, Brüel and Kjær North America, Norcross, GA, USA), driven by an amplified sinusoidal signal from a function generator (Agilent 33250A, Agilent Technologies, Inc. Santa Clara, CA) and an amplifier (Crown XLS202, Crown Audio, Inc. Elkhart, IN), was used to generate a continuous mechanical vibration at a frequency of 2 Hz on the phantom surface during data acquisition [49].

2. *In vivo* studies of human kidney and liver

The *in vivo* studies were approved by the Institutional Review Board of Mayo Clinic, and written informed consent was obtained from the participant. The identical Verasonics vantage system and GE 9L linear array probe were used to acquire kidney and liver data from a volunteer. A 16-angle plane wave compounding imaging was performed (angle increment = 1°, center frequency = 6.25 MHz, pulse length = 2 cycles) and the post-compounded frame rate was 500 Hz for kidney data and 400 Hz for liver data acquisition. A stack of 496 frames of ultrasound IQ data with data type of complex double was captured from the kidney (data size: 180×180×496, pixel size: 0.246 mm) and 600 frames for the liver (data size: 177×173×600, pixel size: 0.256 mm) with the same beamformed pixel size and using the same transmit voltage of 50 V (one-side voltage) as in the phantom study.

3. Data processing and evaluation

All the data were processed offline using MATLAB (version 2021a, MathWorks Inc., Natick, MA, USA) on a PC workstation with a 12-core CPU (Intel® Xeon® W-2265 CPU @ 3.50GHz) and a NVIDIA RTX A6000 GPU. Targeting at the goal of real-time implementation, a relatively short packet size of 50 frames was used for tissue clutter filtering and imaging throughout the study, unless otherwise specified. The data were processed in a sliding temporal window manner, with an interval of 5 frames between two successive packets. The data was first converted to complex single precision data type and transferred to GPU for computational efficiency [50]. The computational time for processing each packet of data to generate a blood flow image was calculated and averaged over all the packets. The processing time included all the processing steps from calculation of covariance matrix (Eqn. 1), transferring the covariance matrix to CPU for EVD computation (Eqn. 2), transferring EVD results back to GPU, automatic eigenvalue thresholding, projection back to the image domain, and image generation.

To quantify the performance of each blood flow image, regions-of-interest (ROIs) of blood flow and background at similar depths will be defined to calculate the contrast ratio ($CR = 10 \log_{10} \overline{PD_{blood}} / \overline{PD_{background}}$) and contrast-to-noise ratio ($CNR = 10 \log_{10} (\overline{PD_{blood}} - \overline{PD_{background}}) / \sigma_{background}$) separately for each data, where $\overline{PD_{blood}}$ and $\overline{PD_{background}}$ are the average intensity of the power Doppler image within the ROIs, respectively, and $\sigma_{background}$ represent the SD within the background ROI. As the data were processed in a sliding window format, the change of CR and CNR over time can be obtained and the mean and SD were calculated. The performance of the adaptive eigenvalue thresholding methods was initially evaluated based on the global data processing setup. The predefined scaling factors for different eigenvalue thresholding methods were exhaustively studied to determine the best achievable performance for each method by maximizing the CR or CNR. The localized clutter filtering method was then implemented, and its performance and computational efficiency were assessed. We also investigated tradeoff between image quality and computational time with changing the packet size.

RESULTS

*Phantom Study*

The performance of the adaptive eigenvalue thresholding methods was first evaluated through global data processing (*i.e.*, cluttering filtering on the full FOV data), as results depicted in Fig. 4. Figure. 4b is an M-mode display of a tissue profile over time to indicate minor tissue motion induced by external vibrator during data acquisition. Clutter filtering was performed packet by packet in a sliding manner along temporal direction (packet size of 50). The mean and SD of CR or CNR were calculated across blood flow images of all packets, for each given scaling factor and each thresholding method. The ROI for CR/CNR quantification is delineated in Fig. 4c. Except for method 4 (minimization of spatial correlation), a predefined scaling factor is required to enable automatic and adaptive determination of eigenvalue cutoff for tissue rejection. Figure. 4d presents CR and CNR as a function of scaling factor for each eigenvalue thresholding method. It is shown that CR or CNR reaches the maximum at a proper scaling factor for each method. A larger or smaller scaler is associated with a reduced CR or CNR, compromising the performance by either inadequate rejection of tissue clutter or excessive removal of flow signal.

The maximum achievable CR and CNR for different methods are illustrated in the left column of Fig. 5d-5e. Among hard thresholding methods, the first three methods (method 1-3) are comparable albeit the CNR for method 3 (the median eigenvalue-based method, 18.8 dB) is slightly higher than the methods 1 and 2 (17.9 dB and 17.8 dB). The computation time of processing one packet of data (from data to the generation of final power Doppler image) is also similar for the first three methods (2.4±1.7 ms, 2.8±0.6 ms and 2.2±1.4 ms, respectively, see left column of Fig. 5f), with method 3 being most computational effective (2.2 ms, corresponding to a processing rate of 455 packet per second). Method 4 is fully automatic based on finding the minimization of spatial correlation without the need for predefining a scaling factor. However, CR and CNR obtained by method 4 are slightly lower than the maximum achievable CR and CNR of other hard thresholding methods (left column of Figs. 5d-5e). The computational cost is also slightly higher for method 4 (3.8±1.0 ms for the global processing on a packet of data, see left column of Fig. 5f), given the requirement to reconstruct the left singular vectors. Method 5 is the soft thresholding based on a sigmoid function centered at the eigenvalue cutoff (provided by method 4), which shows a slightly increased CR and CNR compared with method 4 but is most computationally costly as being built on the top of method 4. The adaptive soft thresholding of method 6 shows the worst CR and CNR performance for this case but requires the minimum computational time (2.0 ms, corresponding to a processing rate of 500 packet per second).

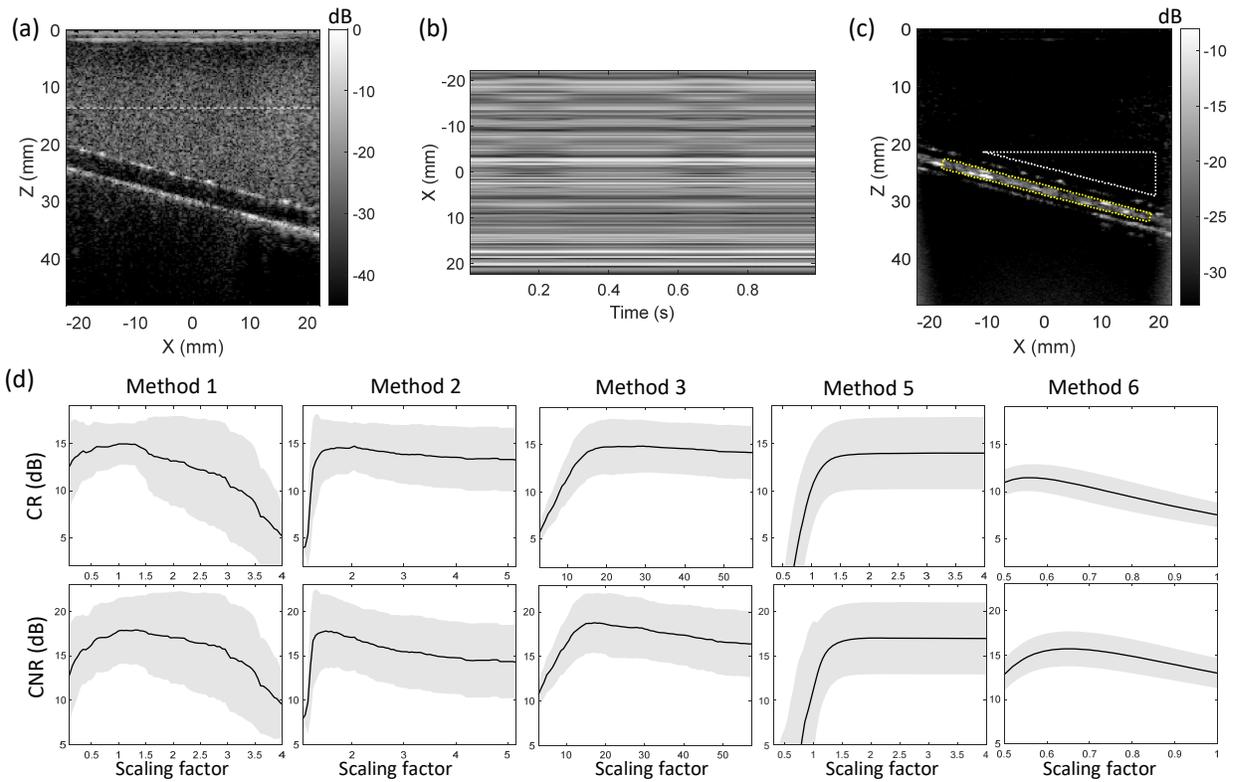

Fig. 4. (a) B-mode image of the flow channel phantom. (b) M-mode representation of the phantom data at profile indicated by the white line in (a). (c) An example of power Doppler image of the flow phantom, with ROIs delineated for CR and CNR computation (white ROI: background; yellow ROI: blood flow). (d) CR and CNR plotted as a function of scaling factor for each adaptive eigenvalue thresholding method based on global data processing. Solid black line represents the mean and gray shadow indicates the range of mean±SD calculated from all data packets.

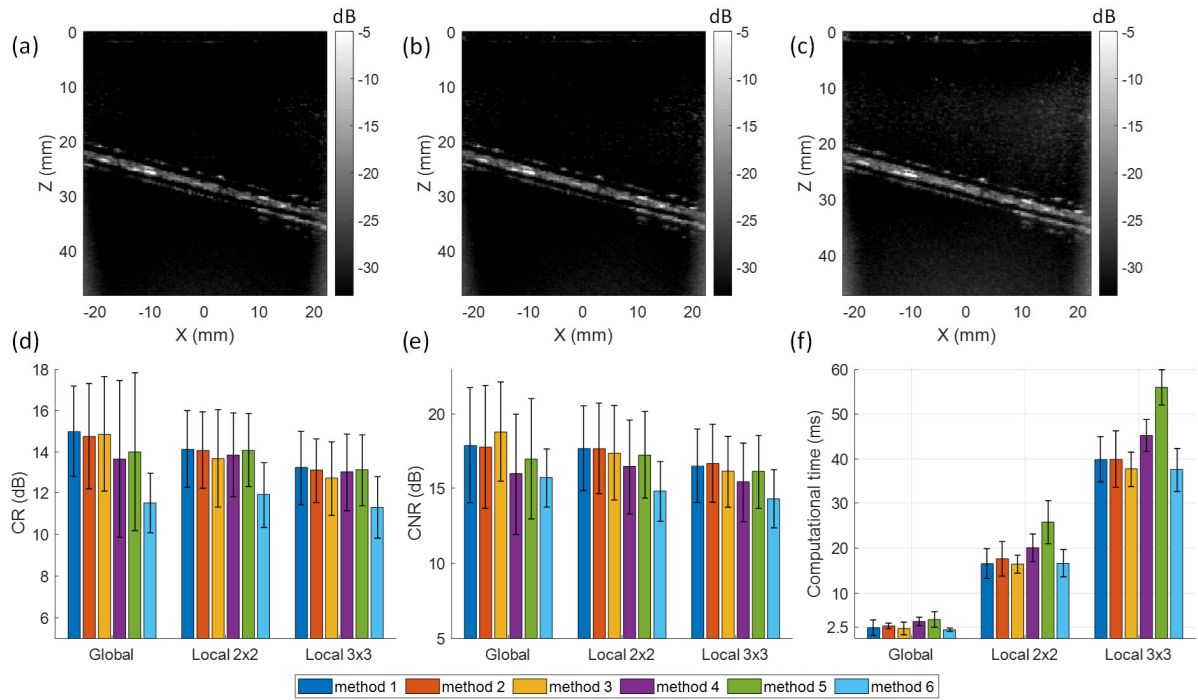

Fig. 5. An example blood flow image of the channel phantom obtained with (a) global processing, (b) 2×2 subsets local processing and (c) 3×3 subsets local processing using thresholding method 1 and a packet size of 50. The maximum achievable (d) CR and (e) CNR for different eigenvalue thresholding methods and for different local processing strategies. (f) The computational time to process one packet of data (packet size of 50) to generate a power Doppler image based on different eigenvalue thresholding methods and local processing strategies.

For localized tissue cluttering, the original data was either divided into 4 subsets (2×2 subsets) or 9 subsets (3×3 subsets), and each subset was sequentially processed and images of subsets were combined to generate the final blood flow image. Example blood flow images obtained by global and local processing are shown in Fig. 5a-5c (packet size = 50, thresholding method 1, scaling factor determined by maximizing CNR). Analogue to the above analysis, all the eigenvalue thresholding methods were evaluated using varying scaling factors, and the maximum achievable CR and CNR are shown in the middle and right column of Fig. 5d-5e. Interestingly, increasing the number of subsets leads to a gradually reduced CR and CNR in this phantom study. This may be attributed to the homogeneous tissue distribution, relatively large vessel, simplified and consistent flow dynamics associated with the phantom study, making it a least challenging scenario for tissue clutter filtering. Additionally, computational time increased with the increased number of subsets, as shown in Fig. 5f. For instance, for tissue clutter filtering using eigenvalue thresholding method 1, the computational time increases from 2.4 ms for global processing, to 16.6 ms and 39.8 ms for the 4 and 9 subsets local processing respectively. Therefore, local processing did not provide performance gains and was less computationally efficient for the current phantom study.

*In Vivo Kidney Study*

The in human experiment provides the most realistic data for the *in vivo* evaluation of methods. The performance of the adaptive eigenvalue thresholding methods based on global processing of the kidney data is presented in Fig. 6. Figure. 6b is the M-mode representation of a tissue profile indicating the small tissue motion over a cardiac cycle, and CR/CNR of all temporal data packets were calculated based on ROIs drawn at a similar depth in Fig. 6c. The CR and CNR are plotted against the changing scaling factor for each eigenvalue thresholding method in Fig. 6d. Again, a proper scaling factor is crucial for enabling automatic and optimal tissue clutter filtering to maximize CR or CNR. It's noteworthy that the scaling factor for maximizing CR may differ from the one for maximizing CNR, owing to the difference in CR and CNR definitions. The left column of Fig. 7d-7e shows the maximum achievable CR and CNR for global clutter filtering using different thresholding methods. Similar levels of maximum CR or CNR can be achieved for the first 4 hard thresholding methods, with less than 1% difference between methods, for both CR and CNR. The computation cost is comparable for the first 3 hard thresholding methods (ranging from 2.5 ms to 3.0 ms, corresponding to processing rate 333 Hz to 400 Hz), with method 3 being the most computational effective. Method 4 is fully automatic with no need to designate a scaling factor, but it requires extra computation times to reconstruct the left singular vectors, thus being the slowest among the hard thresholding filtering methods (3.9 ms). The soft thresholding methods present diverging filtering performances, with substantial increased CR and CNR obtained by method 6 while no significant gain by method 5 for the current kidney data. The computational time is among the lowest for the method 6 (2.5 ms, Fig. 7f).

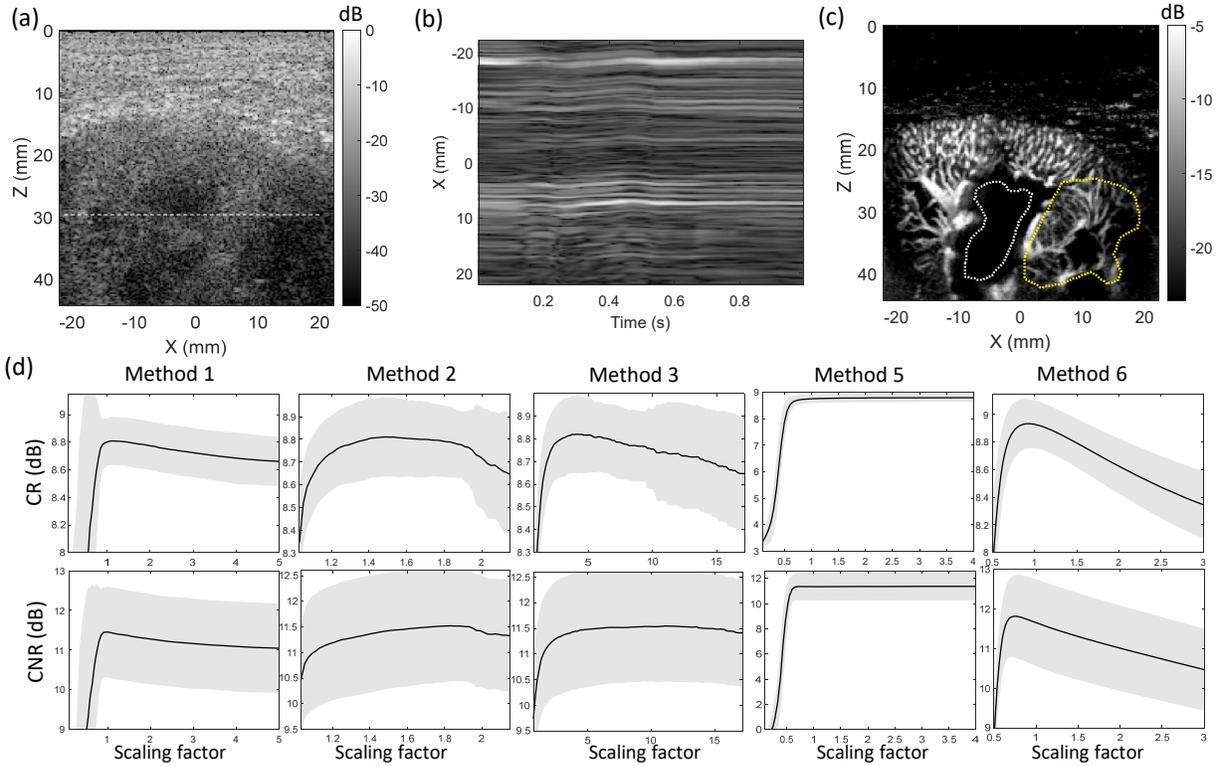

Fig. 6. (a) B-mode image of the *in vivo* kidney. (b) M-mode representation of a tissue profile indicated by the white line in (a). (c) An example of power Doppler image, with ROIs delineated at a similar depth for CR and CNR calculation (white ROI: background; yellow ROI: blood flow). (d) CR and CNR plotted against changing scaling factor for each adaptive eigenvalue thresholding method based on global data processing. Solid black line represents the mean and gray shadow indicates the range of mean±SD calculated from all Doppler data packets.

Different from the phantom data, the *in vivo* data demonstrated significant improved blood flow detection with the increase of local data subsets in local clutter filtering (Fig. 7). Compared with the global processing (Fig. 7a), improved visualization can be observed by the local processing for the weaker flow signals (Figs. 7b-7c, examples indicated by white arrows). The maximum achievable CR and CNR consistently increased with the number of local subsets (middle and right column of Figs. 7d-7e), albeit at a cost of computational time (Fig. 7f). For instance, the CR and CNR increased from 8.8 dB and 11.5 dB with global clutter filtering to 9.7 dB and 12.3 dB with 2×2 subsets local processing, and further to 10.7 dB and 14.1 dB with 3×3 subsets local processing using eigenvalue thresholding method 1. However, the computational time also increased from 2.6 ms for global processing, to 15.7 ms and 36.5 ms for the 2×2 and 3×3 subsets local processing, respectively. For the current kidney data, adaptive soft thresholding method (method 6) achieves the highest CR and CNR, and methods 1, 3 and 6 are among the methods with lowest computational cost.

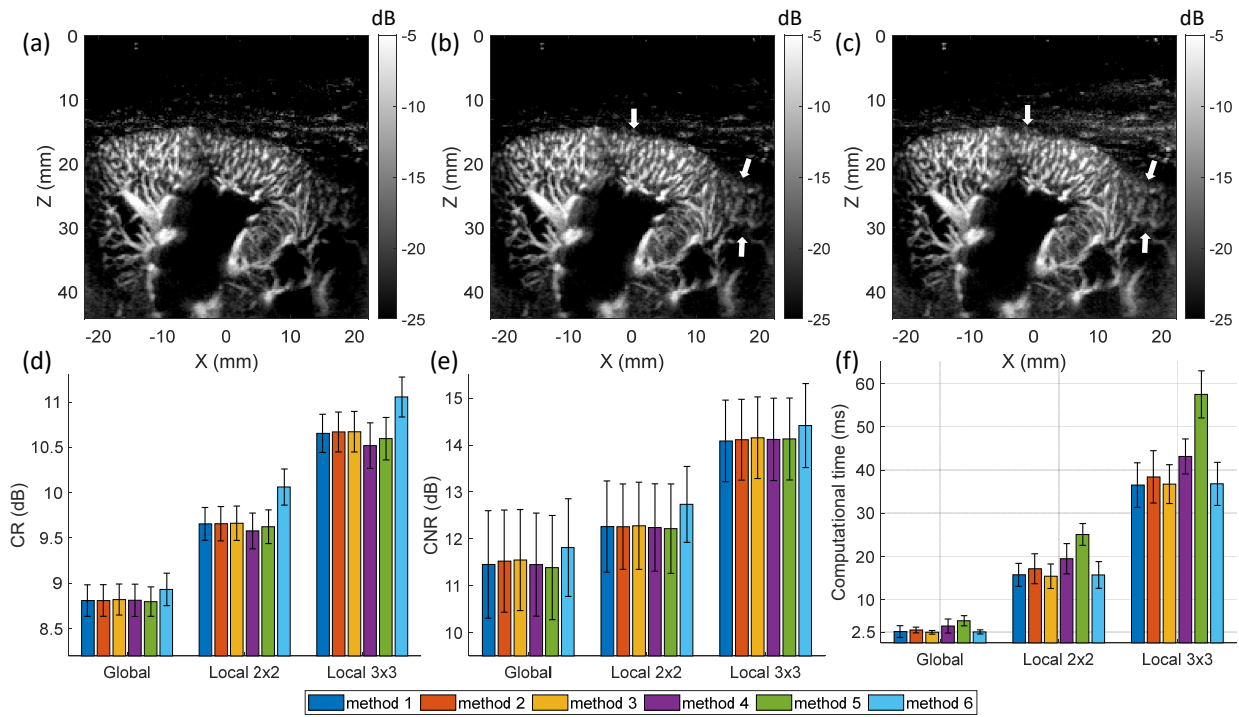

Fig. 7. An example kidney blood flow image obtained with (a) global processing, (b) 2×2 and (c) 3×3 subsets (Supplemental Video 1) local processing using thresholding method 1 at optimal scaling factor and a packet size of 50. White arrows indicate example regions of visually improved vessel detection compared with global processing. The maximum achievable (d) CR, (e) CNR and (f) computational time to process one packet of data (packet size of 50) for different eigenvalue thresholding methods and for different local processing strategies.

*In Vivo Liver Study*

The results of *in vivo* liver blood flow imaging are presented in Figs. 8-9. Analogue to the preceding analysis, CR and CNR were quantified based on the manually drawn ROIs at a similar depth (Fig. 8c), and the adaptive eigenvalue thresholding methods were applied to the liver data with changing scaling factors based on the global clutter filtering (Fig. 8d). Similarly, except for the method 4 that does not require a predefined scaling factor, an appropriate scaling factor is critically to achieve the optimal clutter filtering performance for maximizing either CR or CNR (Fig. 8d). The maximum achievable CR and CNR for different thresholding methods obtained from Fig. 8d are summarized in the left column of Figs. 9d-9e. The SD of the CR and CNR measurements is relatively high in this case, likely due to the large variation of pulsatile blood flow over the cardiac cycle (see Supplemental Video 2). Despite this variability, the maximum CR or CNR tends to be similar among the first 5 thresholding methods (< 5% difference between methods for both CR and CNR), while the method 6 provides the poorest performance. The computational cost results are consistent with those of the kidney data, with the first 3 hard thresholding methods and soft thresholding method 6 requiring lower computational time (ranging from 2.4 ms to 2.9 ms, corresponding to processing rate 345 Hz to 417 Hz) than others.

For the liver data, local processing improves the detection and visualization of small vessels that may be less discernible in global processing, as shown in Figs. 9a-9c (examples indicated by the white arrows). The quantification of CR and CNR is consistent with the visual observation, showing a steady and statistically significant increase of CR and CNR with increased number of local subsets, albeit with a relatively large SD of the measurements (Figs. 9d-9e). For example, for clutter filtering based on thresholding method 1, the CR and CNR were 11.9 dB and 16.3 dB for global processing, which increased to 12.4 dB and 17.0 dB for local processing with 2×2 subsets, and 13.5 dB and 18.1 dB with 3×3 subsets respectively. Corresponding computational time increased from 2.5 ms, to 15.6 ms and 35.2 ms for the 2×2 and 3×3 subsets local processing, respectively. For the current liver data, the eigenvalue thresholding methods 1 and 2 yield the highest achievable CR and CNR, while methods 3 and 6 achieve the lowest computation time, for both global or local processing (Figs. 9d-9f).

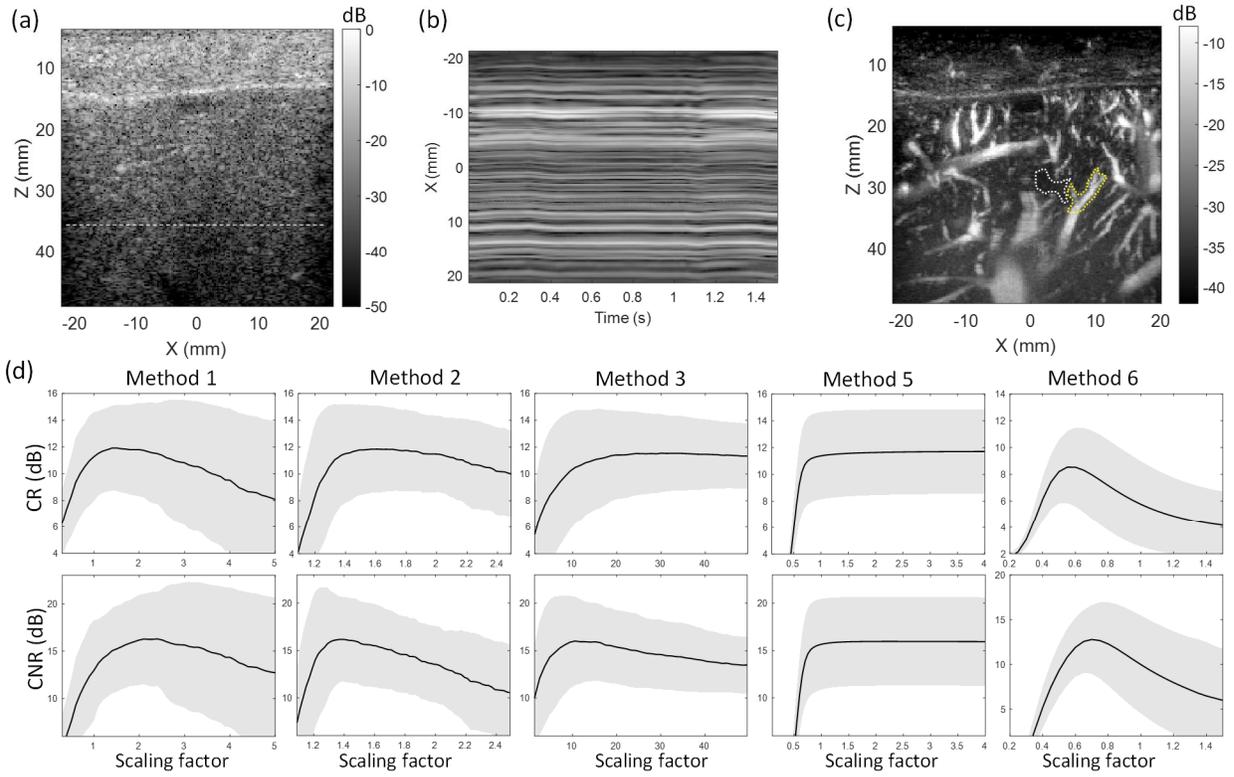

Fig. 8. (a) B-mode image of the *in vivo* liver. (b) M-mode representation of a tissue profile indicated by the white line in (a). (c) An example of the liver power Doppler image, with ROIs draw at a similar depth for CR and CNR calculation (white ROI: background; yellow ROI: blood flow). (d) CR and CNR plotted against scaling factor for each adaptive eigenvalue thresholding method based on global clutter filtering. Solid black line represents the mean and gray shadow indicates the range of mean±SD calculated from all Doppler data packets along temporal direction.

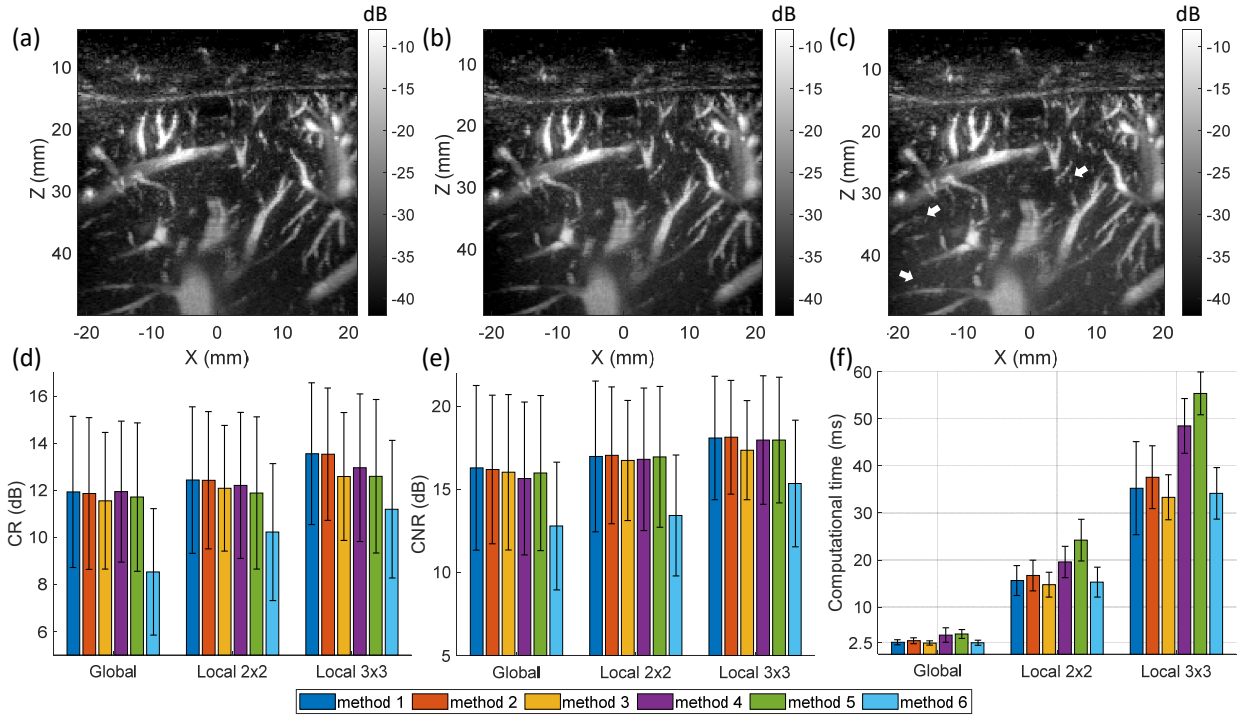

Fig. 9. An example liver blood flow image obtained with (a) global processing, (b) 2×2 and (c) 3×3 subsets local processing using thresholding method 1 (packet size of 50, optimal scaling factor by maximizing CNR). White arrows indicate examples of the small vessels with improved detection compared with global processing. The maximum achievable (d) CR, (e) CNR and (f) computational time for different eigenvalue thresholding methods and different local processing strategies.

*Packet Size Considerations*

We investigated the influence of packet size on blood flow imaging based on the *in vivo* liver data by increasing the packet size from 50 to 70 and 90, as depicted in Fig. 10. A longer packet size was found to improve the detection of small vessels, as finer vessel structures became more discernible in blood flow images with larger packet sizes (Figs. 10a-10c, see Supplemental Videos 2-4). For both global or local processing strategies, an increase of packet size was associated with a consistent increase of CR and CNR, as shown in Figs. 10d-10e. A larger computation cost was expected for both global and local data processing, as shown in Fig. 10f (eigenvalue thresholding method 1 was used here for demonstration). For instance, the computation time of global processing increased from 2.5 ms for packet size of 50 to 2.6 ms and 2.9 ms for the packet size of 70 and 90, corresponding to 4% and 16% increase of computational time, respectively.

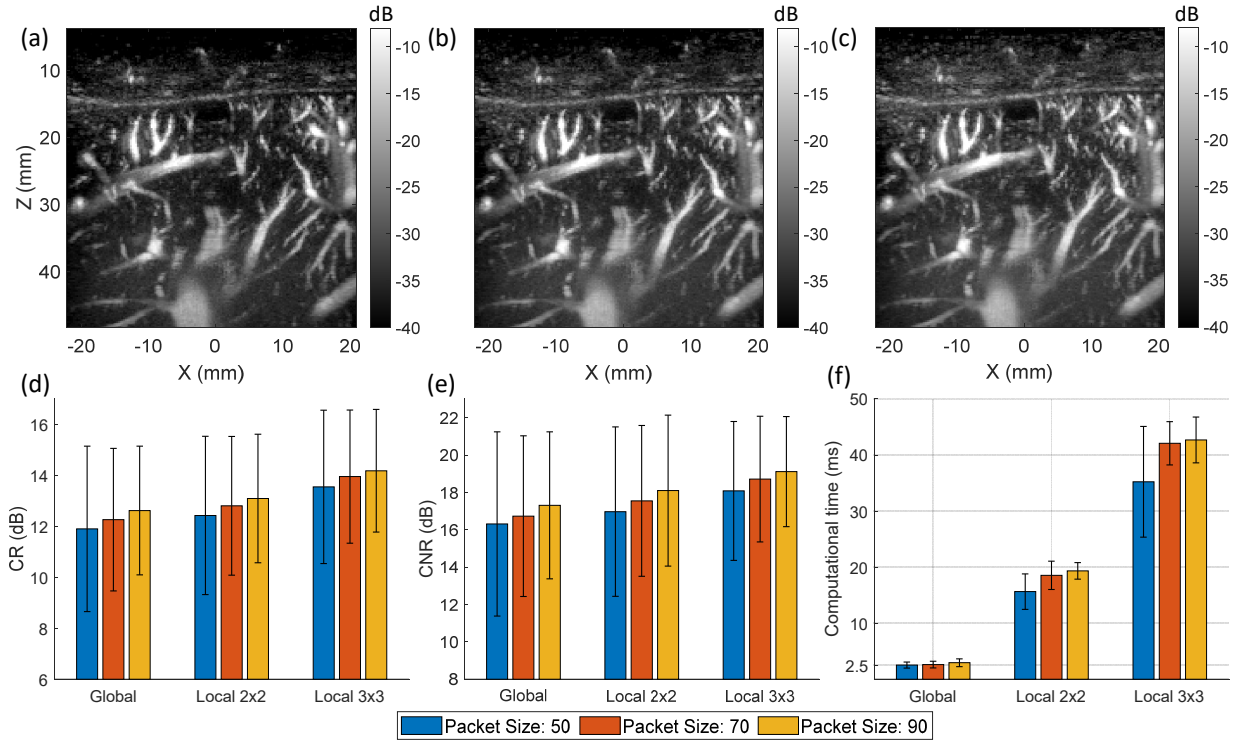

Fig. 10. The example liver blood flow image obtained with packet size of (a) 50, (b) 70 and (c) 90 using local processing (3×3 subsets) and eigenvalue thresholding method 1 (Supplemental Video 2-4). Improved detection of smaller vessels is visually shown with larger packet size. The maximum achievable (d) CR, (e) CNR and (f) computational time for different packet size and different local processing strategies.

## Discussion

In this study, we proposed and evaluated a fast spatiotemporal clutter filtering technique based on EVD and localized data processing for robust and high-sensitive ultrasound imaging of blood flow. We showcase the capability of real-time implementation of the proposed microvessel imaging by leveraging GPU acceleration. The full EVD calculation in current setup enables the adaptive and even fully automatic eigenvalue thresholding based on the characteristics of full eigen spectrum, which allows for robust tissue clutter suppression for broad applications in real-time. We systematically studied several automatic and adaptive eigenvalue thresholding methods, including both hard and soft thresholding, to guide optimization of tissue clutter filtering for both global and localized processing strategies. We further demonstrated the advantages of localized clutter filtering using proposed Gaussian-based spatial windowing strategy *in vivo*, which enhances the sensitivity to small vessels with localized processing while mitigating grid artifacts. A tradeoff between the performance gain by the localized processing and the computational cost was also shown through different *in vivo* applications, including kidney and liver microvessel imaging.

The adaptive identification of tissue components for suppression in the eigen domain is a critical aspect of successful tissue clutter filtering, particularly in dynamic *in vivo* imaging scenarios where tissue motion and flow dynamics vary over the cardiac cycle. Additionally, localized tissue clutter filtering presents challenges due to spatially varying data characteristics, necessitating robust identification methods adaptive to the different characteristics associated with different local data subsets. In most eigenvalue thresholding methods for tissue clutter filtering, a predefined scaling factor is required to initiate the adaptive and automatic identification of tissue or blood components. For example, a proper slope of the eigenvalue curve is predefined as the scaling factor to enable the identification of eigen components with slopes above this scaling factor to be thresholded. Among the methods studied in this paper, the first 3 hard eigenvalue thresholding methods (based on curvature, slope, or median eigenvalue) demonstrated similar performance when a proper scaling factor was provided. However, a different scaling factor will be needed for optimal filtering for different applications and different imaging setups (such as data size and packet size). Interestingly, the optimal scaling factor for methods 1 and 2 showed less divergence (falling in a range of 1~2) across different applications (including phantom, kidney, and liver), implying that less intervention may be required to determine a suitable scaling factor for broader applications. On the other hand, the range of optimal scaling factor for method 3 was wider, despite its minimal computational cost among hard thresholding methods. Method 4 does not require a predefined scaling factor and enables a fully automatic tissue clutter filtering based on the assumption that tissue and blood flow are most distinct in spatial structure at their best separation. However, the achievable CR and CNR are slightly lower than the other methods. In contrast, soft thresholding methods exhibit highly divergent performance across applications, with improved performance for some applications (*e.g.*, kidney) and decreased performance for others (*e.g.*, liver). As the eigen components are weighted based on their corresponding eigenvalues in soft thresholding, the diverging performance may be highly related to the eigenvalue distribution in the eigen spectrum, which varies from case to case. Another significant limitation of soft thresholding is that the relative contribution of different blood eigen components has been modified, potentially leading to dominance of certain components over others in the final blood flow image. Therefore, although computational effective, soft thresholding methods should be used and interpreted with caution.

Local tissue cluttering leverages the unique characteristics of local data to enhance the sensitivity of small vessel detection *in vivo* at a cost of computational burden. However, the benefits of this local processing strategy may not always be applicable to simple blood flow imaging scenarios where tissue characteristics are spatially homogenous across the FOV, such as the case of phantom imaging presented in this study. In general, a larger FOV provides richer spatial information that may contribute to EVD calculation for better differentiation of tissue and blood in eigen domain, assuming consistent tissue and noise statistics across the entire FOV. Therefore, for the flow phantom imaging where tissue is mostly homogenous, and the flow channel is relatively large and straight with simplified flow dynamics, a global processing of the large FOV data is expected to improve the clutter filtering. Consequently, local processing strategy may not be necessary in such cases with less heterogeneous distribution of tissue, blood flow, and noise, or even in *in vivo* cases when the FOV is small. With the proposed EVD-based adaptive clutter filtering framework, global processing is most computationally efficient, achieving processing rate of up to 500 Hz (packet size of 50) while maintaining a promising performance. Nevertheless, in complex *in vivo* applications where significant spatial heterogeneity is present across a large FOV, local processing is recommended to further boost the microvessel detection, albeit with an associated increase in computational time. The proposed local processing strategy allows for the use of a small number of data subsets to achieve localized filtering while mitigating potential grid artifacts. Despite the additional computational expense associated with these limited subsets, advancements in computational hardware, such as GPU, make real-time implementation still affordable. For example, in this study, processing rates of about 60 Hz for a 2x2, and about 25 Hz for a 3×3 local processing setup (packet size of 50 frames) were attainable for the current *in vivo* data. One may choose an appropriate number of subsets in local processing to balance the trade-off between computational efficiency and performance considering the available hardware and specific clinical application requirements.

A larger packet size offers advantages for improving microvessel detection, as illustrated in this study, but at the expense of both computational and acquisition time. The benefits of a longer packet size are twofold: firstly, a longer packet size provides more spatial-temporal information for better separation of tissue and blood flow in eigen domain, thereby improving detection of slow flow signal; and secondly, it allows for data accumulation over a longer duration to average out the noisy background and improve sensitivity to weak flow signal [7]. While theoretical computational complexity of principal component analysis is $O(N_xN_yN_t^2+N_t^3)$ [47], in practice the increase in computational time from a packet size of 50 to 90 (80% increase of $N_t$) is relatively minor (less than 20% increase of computation time) in this study. This efficiently may be attributed in part to the parallel computation capacity of GPUs for accelerated calculation of the covariance matrix, which represents the primary computation burden when $N_xN_y$ is much larger than $N_t$. However, besides the computational cost of clutter filtering, factors such as data acquisition time (a packet size of 50 at a PRF of 500 Hz corresponding to a date length of 0.1 s), data transferring between device and host, beamforming and computational overhead also need to be considered in the entire data generation and processing pipeline for real-time implementation. As demonstrated in this study, tissue clutter filtering may no longer be a significant computational bottleneck (processing up to 500 packets per second using global processing), a sliding window processing strategy that implements clutter filtering more frequently on packets with certain temporal overlaps can be employed to enable a higher frame rate display of microvessel image in the real-time implementation. Nevertheless, a shorter packet is still preferred for achieving higher temporal resolution of microvessel imaging.

This study has several limitations that warrant consideration. Firstly, the optimal performance of different eigenvalue thresholding methods (except for method 4) was obtained by implementing tissue clutter filtering under various predefined scaling factors, which is impractical for real-time implementation. For clinical application, an optimal scaling factor will still need to be empirically determined based on the imaging setups (*e.g.*, FOV, packet size, PRF) and specific clinical scenarios. For real-time implementation, a user interface is recommended to allow the user to fine-tune the scaling factor or eigenvalue cutoff to achieve an optimal imaging performance in clinical practice. Moreover, the quantitative CR and CNR was used as the criteria to define the optimal imaging performance and the corresponding optimal scaling factors. However, defining optimal performance can be ambiguous, and the optimal scaling factor can be varying depending on the desired imaging outcome. There may also be a discrepancy between 'optimal performance' based on quantitative criteria and subjective visual inspection, as human observers often prefer visually cleaner images with clutters excessively suppressed. In this case, a user interface allowing user input to fine-tune the clutter filtering may be beneficial. Additionally, the current localized tissue clutter filtering approach processes each subset sequentially in a for-loop, leading to computational time proportional to the number of subsets. A parallel processing of all data subsets instead of sequential processing may further enhance computational efficiency for local tissue clutter filtering, which remains for future investigation. In this study, we focused on evaluating relatively small packet sizes (50~90) with the goal for real-time implementation. However, an even higher definition version of microvessel imaging could be implemented based on longer data packets, preferably captured during patient breath-hold, with acceptable extra processing time. Another consideration is the use of single precision data type in our current study to conserve memory and maximize computation efficiency for real-time microvessel imaging. While single precision is in general sufficient for processing short data packets without significant compromise in imaging performance, we found that the computational accuracy of EVD may be hampered when handling a larger dataset, such as data with much longer packet size. In such cases, a double precision may be essential to ensure robust tissue clutter filtering.

In conclusion, we developed and tested a fast spatiotemporal clutter filtering technique based on EVD and novel localized data processing for fully adaptive and highly sensitive ultrasound imaging of small vessels. We demonstrated the real-time capability of our proposed approach by harnessing the computational power of GPU. The feasibility of the proposed technique was validated in different *in vivo* applications in this study, including kidney and liver, highlighting the potential of the approach for a broad clinical translation.

REFERENCES

[1] S. Bjaerum, H. Torp, and K. Kristoffersen, "Clutter filter design for ultrasound color flow imaging," *IEEE Transactions on Ultrasonics, Ferroelectrics, and Frequency Control,* vol. 49, no. 2, pp. 204-216, 2002.

[2] D. H. Evans, J. A. Jensen, and M. B. Nielsen, "Ultrasonic colour Doppler imaging," *Interface Focus,* vol. 1, no. 4, pp. 490-502, Aug 6 2011.

[3] H. Torp, "Clutter rejection filters in color flow imaging: a theoretical approach," *IEEE Trans Ultrason Ferroelectr Freq Control,* vol. 44, no. 2, pp. 417-24, 1997.

[4] A. C. H. Yu and L. Lovstakken, "Eigen-Based Clutter Filter Design for Ultrasound Color Flow Imaging: A Review," *IEEE Trans. Ultrason. Ferroelectr. Freq. Control,* vol. 57, no. 5, pp. 1096-1111, May 2010.

[5] C. Demene *et al.*, "Spatiotemporal Clutter Filtering of Ultrafast Ultrasound Data Highly Increases Doppler and fUltrasound Sensitivity," *IEEE Trans. Med. Imaging,* vol. 34, no. 11, pp. 2271-2285, Nov 2015.

[6] J. Bercoff *et al.*, "Ultrafast Compound Doppler Imaging: Providing Full Blood Flow Characterization," *IEEE Trans. Ultrason. Ferroelectr. Freq. Control,* vol. 58, no. 1, pp. 134-147, Jan 2011.

[7] E. Mace, G. Montaldo, B. F. Osmanski, I. Cohen, M. Fink, and M. Tanter, "Functional Ultrasound Imaging of the Brain: Theory and Basic Principles," *IEEE Trans. Ultrason. Ferroelectr. Freq. Control,* vol. 60, no. 3, pp. 492-506, Mar 2013.

[8] S. Bjaerum, H. Torp, and K. Kristoffersen, "Clutter filters adapted to tissue motion in ultrasound color flow imaging," *IEEE Trans Ultrason Ferroelectr Freq Control,* vol. 49, no. 6, pp. 693-704, Jun 2002.

[9] L. A. Ledoux, P. J. Brands, and A. P. Hoeks, "Reduction of the clutter component in Doppler ultrasound signals based on singular value decomposition: a simulation study," *Ultrason Imaging,* vol. 19, no. 1, pp. 1-18, Jan 1997.

[10]  L. Løvstakken, S. Bjaerum, K. Kristoffersen, R. Haaverstad, and H. Torp, "Real-time adaptive clutter rejection filtering in color flow imaging using power method iterations," *IEEE Trans Ultrason Ferroelectr Freq Control,* vol. 53, no. 9, pp. 1597-608, Sep 2006.

[11]  D. E. Kruse and K. W. Ferrara, "A new high resolution color flow system using an eigendecomposition-based adaptive filter for clutter rejection," *IEEE Transactions on Ultrasonics, Ferroelectrics, and Frequency Control,* vol. 49, no. 10, pp. 1384-1399, 2002.

[12]  F. Song, D. Zhang, and X. Gong, "Performance evaluation of eigendecomposition-based adaptive clutter filter for color flow imaging," *Ultrasonics,* vol. 44 Suppl 1, pp. e67-71, Dec 22 2006.

[13]  C. M. Gallippi and G. E. Trahey, "Adaptive clutter filtering via blind source separation for two-dimensional ultrasonic blood velocity measurement," *Ultrason Imaging,* vol. 24, no. 4, pp. 193-214, Oct 2002.

[14]  A. C. Yu and R. S. Cobbold, "Single-ensemble-based eigen-processing methods for color flow imaging--Part I. The Hankel-SVD filter," *IEEE Trans Ultrason Ferroelectr Freq Control,* vol. 55, no. 3, pp. 559-72, Mar 2008.

[15]  A. J. Y. Chee, B. Y. S. Yiu, and A. C. H. Yu, "A GPU-Parallelized Eigen-Based Clutter Filter Framework for Ultrasound Color Flow Imaging," *IEEE Transactions on Ultrasonics, Ferroelectrics, and Frequency Control,* vol. 64, no. 1, pp. 150-163, 2017.

[16]  W. You and Y. Wang, "A fast algorithm for adaptive clutter rejection in ultrasound color flow imaging based on the first-order perturbation: a simulation study," *IEEE Trans Ultrason Ferroelectr Freq Control,* vol. 57, no. 8, pp. 1884-9, Aug 2010.

[17]  M. Tanter, J. Bercoff, L. Sandrin, and M. Fink, "Ultrafast compound imaging for 2-D motion vector estimation: Application to transient elastography," *IEEE Trans. Ultrason. Ferroelectr. Freq. Control,* vol. 49, no. 10, pp. 1363-1374, Oct 2002.

[18]  G. Montaldo, M. Tanter, J. Bercoff, N. Benech, and M. Fink, "Coherent Plane-Wave Compounding for Very High Frame Rate Ultrasonography and Transient Elastography," *IEEE Trans. Ultrason. Ferroelectr. Freq. Control,* vol. 56, no. 3, pp. 489-506, Mar 2009.

[19] C. Huang *et al.*, "Noninvasive Contrast-Free 3D Evaluation of Tumor Angiogenesis with Ultrasensitive Ultrasound Microvessel Imaging," *Sci Rep,* vol. 9, no. 1, p. 4907, 2019/03/20 2019.

[20] P. F. Song *et al.*, "Functional Ultrasound Imaging of Spinal Cord Hemodynamic Responses to Epidural Electrical Stimulation: A Feasibility Study," *Front. Neurol.,* vol. 10, p. 13, Mar 2019, Art. no. 279.

[21] W. Zhang *et al.*, "Ultrasensitive US Microvessel Imaging of Hepatic Microcirculation in the Cirrhotic Rat Liver," *Radiology,* vol. 307, no. 1, p. e220739, Apr 2023.

[22] P. Gong *et al.*, "Quantitative Inflammation Assessment for Crohn Disease Using Ultrasensitive Ultrasound Microvessel Imaging: A Pilot Study," *J Ultrasound Med,* vol. 39, no. 9, pp. 1819-1827, Sep 2020.

[23] R. R. Wildeboer *et al.*, "Blind Source Separation for Clutter and Noise Suppression in Ultrasound Imaging: Review for Different Applications," *IEEE Transactions on Ultrasonics, Ferroelectrics, and Frequency Control,* vol. 67, no. 8, pp. 1497-1512, 2020.

[24] F. W. Mauldin, Jr., D. Lin, and J. A. Hossack, "The singular value filter: a general filter design strategy for PCA-based signal separation in medical ultrasound imaging," *IEEE Trans Med Imaging,* vol. 30, no. 11, pp. 1951-64, Nov 2011.

[25] K. Riemer *et al.*, "On the use of singular value decomposition as a clutter filter for ultrasound flow imaging," *arXiv preprint arXiv:2304.12783,* 2023.

[26] J. Tierney, J. Baker, D. Brown, D. Wilkes, and B. Byram, "Independent Component-Based Spatiotemporal Clutter Filtering for Slow Flow Ultrasound," *IEEE Trans. Med. Imaging,* vol. 39, no. 5, pp. 1472-1482, 2020.

[27] K. A. Ozgun and B. C. Byram, "Multidimensional Clutter Filtering of Aperture Domain Data for Improved Blood Flow Sensitivity," *IEEE Trans Ultrason Ferroelectr Freq Control,* vol. 68, no. 8, pp. 2645-2656, Aug 2021.

[28] M. Kim, Y. Zhu, J. Hedhli, L. W. Dobrucki, and M. F. Insana, "Multidimensional Clutter Filter Optimization for Ultrasonic Perfusion Imaging," *IEEE Trans. Ultrason. Ferroelectr. Freq. Control,* vol. 65, no. 11, pp. 2020-2029, Nov 2018.

[29] M. Kim, C. K. Abbey, J. Hedhli, L. W. Dobrucki, and M. F. Insana, "Expanding Acquisition and Clutter Filter Dimensions for Improved Perfusion Sensitivity," *IEEE Trans Ultrason Ferroelectr Freq Control,* vol. 64, no. 10, pp. 1429-1438, Oct 2017.

[30] Y. Sui *et al.*, "Randomized Spatial Downsampling-Based Cauchy-RPCA Clutter Filtering for High-Resolution Ultrafast Ultrasound Microvasculature Imaging and Functional Imaging," *IEEE Trans Ultrason Ferroelectr Freq Control,* vol. 69, no. 8, pp. 2425-2436, Aug 2022.

[31] D. H. Pham, A. Basarab, I. Zemmoura, J. P. Remenieras, and D. Kouame, "Joint Blind Deconvolution and Robust Principal Component Analysis for Blood Flow Estimation in Medical Ultrasound Imaging," *IEEE Trans Ultrason Ferroelectr Freq Control,* vol. 68, no. 4, pp. 969-978, Apr 2021.

[32] N. Zhang, M. Ashikuzzaman, and H. Rivaz, "Clutter suppression in ultrasound: performance evaluation and review of low-rank and sparse matrix decomposition methods," *Biomed Eng Online,* vol. 19, no. 1, p. 37, May 28 2020.

[33] O. Solomon *et al.*, "Deep Unfolded Robust PCA With Application to Clutter Suppression in Ultrasound," *IEEE Trans. Med. Imaging,* vol. 39, no. 4, pp. 1051-1063, 2020.

[34] P. F. Song, A. Manduca, J. D. Trzasko, and S. G. Chen, "Ultrasound Small Vessel Imaging With Block-Wise Adaptive Local Clutter Filtering," *IEEE Trans. Med. Imaging,* vol. 36, no. 1, pp. 251-262, Jan 2017.

[35] P. F. Song *et al.*, "Accelerated Singular Value-Based Ultrasound Blood Flow Clutter Filtering With Randomized Singular Value Decomposition and Randomized Spatial Downsampling," *IEEE Trans. Ultrason. Ferroelectr. Freq. Control,* vol. 64, no. 4, pp. 706-716, Apr 2017.

[36] U. W. Lok *et al.*, "Real time SVD-based clutter filtering using randomized singular value decomposition and spatial downsampling for micro-vessel imaging on a Verasonics ultrasound system," *Ultrasonics,* vol. 107, p. 106163, 2020/09/01/ 2020.

[37] B. Pialot, L. Augeul, L. Petrusca, and F. Varray, "A simplified and accelerated implementation of SVD for filtering ultrafast power Doppler images," *Ultrasonics,* vol. 134, p. 107099, Sep 2023.

[38] J. Baranger, B. Arnal, F. Perren, O. Baud, M. Tanter, and C. Demene, "Adaptive spatiotemporal SVD clutter filtering for Ultrafast Doppler Imaging using similarity of spatial singular vectors," *IEEE Trans. Med. Imaging,* vol. PP, no. 99, pp. 1-1, Feb 2018.

[39] D. Maresca, M. Correia, M. Tanter, B. Ghaleh, and M. Pernot, "Adaptive Spatiotemporal Filtering for Coronary Ultrafast Doppler Angiography," *IEEE Trans Ultrason Ferroelectr Freq Control,* vol. 65, no. 11, pp. 2201-2204, Nov 2018.

[40] S. A. Waraich, A. Chee, D. Xiao, B. Y. S. Yiu, and A. Yu, "Auto SVD Clutter Filtering for US Doppler Imaging Using 3D Clustering Algorithm," in *Image Analysis and Recognition*, Cham, 2019, pp. 473-483: Springer International Publishing.

[41] B. Arnal, J. Baranger, C. Demene, M. Tanter, and M. Pernot, "In vivo real-time cavitation imaging in moving organs," *Phys Med Biol,* vol. 62, no. 3, pp. 843-857, Feb 7 2017.

[42] J. Baranger, J. Aguet, and O. Villemain, "Fast Thresholding of SVD Clutter Filter Using the Spatial Similarity Matrix and a Sum-Table Algorithm," *IEEE Trans Ultrason Ferroelectr Freq Control,* vol. 70, no. 8, pp. 821-830, Aug 2023.

[43] Y. Chen, B. Fang, F. Meng, J. Luo, and X. Luo, "Competitive Swarm Optimized SVD Clutter Filtering for Ultrafast Power Doppler Imaging," *IEEE Transactions on Ultrasonics, Ferroelectrics, and Frequency Control,* vol. 71, no. 4, pp. 459-473, 2024.

[44] H. Abdi and L. J. Williams, "Principal component analysis," *WIREs Computational Statistics,* vol. 2, no. 4, pp. 433-459, 2010.

[45] T. Dobravec and P. Bulić, "Comparing CPU and GPU Implementations of a Simple Matrix Multiplication Algorithm," *International Journal of Computer and Electrical Engineering,* vol. 9, pp. 430-438, 2017.

[46] D. Guide, "Cuda c best practices guide," *NVIDIA, July,* 2013.

[47] N. Kishore Kumar and J. Schneider, "Literature survey on low rank approximation of matrices," *Linear and Multilinear Algebra,* vol. 65, no. 11, pp. 2212-2244, 2017/11/02 2017.

[48] M. Gavish and D. L. Donoho, "The Optimal Hard Threshold for Singular Values is $4/\sqrt{3}$," *IEEE Transactions on Information Theory,* vol. 60, no. 8, pp. 5040-5053, 2014.

[49] C. Huang *et al.*, "Three-dimensional shear wave elastography on conventional ultrasound scanners with external vibration," *Phys Med Biol,* vol. 65, no. 21, p. 215009, Nov 5 2020.

[50] N. J. Higham and T. Mary, "Mixed precision algorithms in numerical linear algebra," *Acta Numerica,* vol. 31, pp. 347-414, 2022.